\documentclass{article}
\usepackage[a4paper, margin=1in]{geometry} %
\usepackage{footnote}
\usepackage{rotating}
\usepackage{graphicx} 
\usepackage{color}
\usepackage{booktabs}
\usepackage{comment}
\usepackage{amsmath}
\usepackage{tabularx}
\usepackage{comment}
\usepackage{hyperref}
\usepackage{caption}
\usepackage{subfigure}
\usepackage{subcaption}
\usepackage{tablefootnote}
\makesavenoteenv{tabular}
\makesavenoteenv{table}
\usepackage{makecell}
\usepackage{authblk}
\hypersetup{
	bookmarks=true,         
	unicode=false,          
	pdftoolbar=true,        
	pdfmenubar=true,        
	pdffitwindow=false,     
	pdfstartview={FitH},    
	pdftitle={My title},    
	pdfauthor={Author},     
	pdfsubject={Subject},   
	pdfcreator={Creator},   
	pdfproducer={Producer}, 
	pdfkeywords={keywords}, 
	pdfnewwindow=true,      
	colorlinks=true,       
	linkcolor=blue,          
	citecolor=blue,        
	filecolor=magenta,      
	urlcolor=blue           
}

\newcommand{\tb}{\textcolor{black}}

\title{Network reciprocity turns cheap talk into a force for cooperation}

\author[1]{Zhao Song\thanks{Z.Song@tees.ac.uk}}
\author[2]{Chen Shen}
\author[1]{The Anh Han}
\affil[1]{School of Computing, Engineering and Digital Technologies, Teesside University, United Kingdom}
\affil[2]{Faculty of Engineering Sciences, Kyushu University, Japan}

\begin{document}

\maketitle

\begin{abstract}
Non-binding communication is common in daily life and crucial for fostering cooperation, even though it has no direct payoff consequences. However, despite robust empirical evidence, its evolutionary basis remains poorly understood. Here, we develop a game-theoretic model in which individuals can signal an intention to cooperate before playing a Donation game. Strategies differ in how they respond to these signals, ranging from unconditional to conditional types, with the latter incurring a cognitive cost for deliberation. Through evolutionary analysis, we show that non-binding communication alone cannot sustain cooperation in well-mixed, anonymous populations, consistent with empirical observations. In contrast, structured populations support the emergence of cooperation, with conditional cooperators acting as catalysts that protect unconditional cooperators through context-dependent patterns of cyclic dominance. These findings offer an evolutionary explanation for how non-binding communication promotes cooperation and provide a modelling framework for exploring its effects in diverse social settings.



\end{abstract}

\section{Introduction}

Humans possess unique communication abilities that enable them to solve complex social problems through information exchange, from negotiating climate risks between nations to coordinating public health responses during epidemics, to resolving everyday conflicts through conversation~\cite{victor2022determining,bavel2020using,smith2010communication}. This capacity has naturally drawn attention in the study of cooperative behaviour, a longstanding puzzle across disciplines ~\cite{dietz2003struggle,boyd2009culture}. Cooperation involves paying a personal cost to benefit others, but is continually threatened by free riders who exploit others’ contributions~\cite{rand2013human,perc2017statistical}.  The Donation game captures this dilemma: individuals choose whether to cooperate, incurring a cost to help a partner, or to defect and maximise their own payoff ~\cite{hamilton1964genetical}. Although mutual cooperation produces the highest collective benefit, defection is the individually optimal strategy, often leading to the breakdown of cooperation.

Economic experiments show that allowing participants to communicate before making decisions increases cooperation, even when communication is non-binding and has no direct payoff consequences ~\cite{crawford1998survey,andersson2012credible,feldhaus2016more}. This effect relies on trust and tends to appear transiently in one-shot, anonymous interactions in well-mixed populations~\cite{duffy2002actions,erbaugh2024communication,charness2006promises}, often fading as interactions proceed. By contrast, when games are repeated among the same individuals, the positive impact of communication on cooperation tends to persist ~\cite{blume2007effects,tingley2011can,sutter2009communication}. Meta-analytic studies confirm this pattern across different communication types and timing relative to the game~\cite{balliet2010communication,sally1995conversation}. Recent work extends these findings to noisy environments, showing that cheap talk promotes cooperation only when players have self-interested motivations to cooperate in repeated interactions~\cite{arechar2017m,dvorak2024negotiating}. Despite the empirical robustness of these results, it remains theoretically unclear why costless, non-enforceable communication should promote cooperation from an evolutionary perspective ~\cite{ellingsen2010does,bahel2022communication,bernheim1987coalition}.

Evolutionary explanations for non-binding communication remain limited, as standard models typically focus on how cooperation evolves among payoff-driven individuals ~\cite{axelrod1981evolution, nowak2006evolutionary}. Since cheap talk does not alter the game’s payoff structure, it has no effect on evolutionary outcomes under conventional assumptions. As a result, most theoretical work has focused on binding communication, where signals are tied to material consequences such as costs for deception or rewards for honesty ~\cite{gintis2001costly,lang2024role,han2016emergence}. These models aim to identify conditions under which cooperation can be evolutionarily stable across diverse contexts~\cite{han2013emergence,salahshour2019evolution,duong2021cost}.

Nevertheless, individuals may differ in how they process and respond to signals, even when the signals carry no direct payoff consequences ~\cite{camerer2004cognitive}. Such variation can stem from underlying cognitive processes, as described by dual-process theories, which distinguish between fast, intuitive strategies and slower, deliberative ones ~\cite{bear2016intuition,rubinstein2007instinctive}. While deliberation can enable more strategic responses, it also imposes cognitive costs. These costs create fitness trade-offs, opening the possibility of modelling responses to non-binding communication as heritable traits subject to evolutionary selection.

To examine this possibility, we develop an evolutionary model that explores the role of non-binding communication in promoting cooperation among self-interested individuals, inspired by dual-process theories of cognition~\cite{bear2016intuition}. The game unfolds in two stages: in the first, players choose whether to signal an intention to cooperate; in the second, they choose to cooperate or defect. We consider four strategies. The two intuitive types are unconditional cooperators ($UC$), who always signal and cooperate, and unconditional defectors ($UD$), who never signal and always defect. The two deliberative types, which incur a cognitive cost, are conditional cooperators ($CC$), who cooperate only if their opponent also signals, and strategic defectors ($CD$), who signal but always defect.

Through evolutionary simulations, we confirm that in well-mixed populations, non-binding communication alone cannot sustain cooperation in one-shot, anonymous interactions. In contrast, in structured populations, cheap talk promotes cooperation through intricate dynamic pathways shaped by the cost of social reasoning. When this cost falls within an intermediate range, cooperation peaks, driven by context-dependent patterns of cyclic dominance. Outside this regime, cooperation may persist through baseline network reciprocity, but otherwise collapses. These results provide an evolutionary explanation for how costless communication can support cooperation in structured populations and offer a general framework for studying cheap talk in diverse social contexts.

\section{Models and Methods}
\subsection{Two-stage game with cheap talk}
We formulate cheap talk within a two-stage game framework, integrating pre-game intention-signalling and in-game decision-making. 
In the pre-game stage, players simultaneously decide to signal a cooperative intention ($S$) or remain silent ($N$).
This communication is ``cheap talk", as the signal itself carries no direct cost and is non-enforceable.
After observing each other's signals, players independently choose to either cooperate ($C$) or defect ($D$). These decisions take place in the context of a social dilemma, which we introduce via the classic Donation game. In the Donation game, players choosing $C$ incur a cost $c$ to provide a benefit $b$ ($b>c$) to the co-player, while players choosing $D$ incur no cost and provide no benefit. Thus, mutual cooperation generates a reward $R=b-c$, mutual defection incurs a punishment $P=0$ for each player, and unilateral actions lead to an asymmetry: a cooperator receives a sucker's payoff $S=-c$, while a defector receives a temptation payoff $T=b$. After rescaling and substituting, the payoffs become $R=1$, $S=-r$, $T=1+r$, and $P=0$, a special case of the Prisoner's Dilemma game ~\cite{axelrod1988further}. Here $r>0$ quantifies the dilemma strength ~\cite{wang2015universal}.

A player's strategy in the game is determined by three components: their action in the first stage ($S$, $N$), their response if the co-player signals ($C$, $D$), and their response if the co-player remains silent ($C$, $D$). This results in eight possible strategies summarized in Table \ref{tab:8strategy}. 
To reduce complexity, our model focuses on four archetypal strategies, reflecting distinct cognitive processes: \textit{intuitive} and \textit{deliberative}. 
\begin{itemize}
    \item $UC$ (unconditional cooperation), where players always signal cooperative intention in the pre-game stage and always cooperate in the game, regardless of co-player's intention; 
    \item $CC$ (conditional cooperation), where players always signal the cooperative intention but only cooperate if the co-player shows cooperative intention as well, otherwise, defect; 
    \item $UD$ (unconditional defection), where players never show cooperative intention and always defect; 
    \item $CD$ (strategic defection), where players pretend a cooperative intention while always defecting to exploit the co-player.
\end{itemize}
These strategies capture a spectrum of human behaviours: $CC$ and $CD$ represent the deliberative and flexible choices, where although cooperation intentions are frequently pronounced before the game, promise breaches occur; whereas $UC$ and $UD$ reflect intuitive and inflexible choices, where the choice aligns with the intention signalled. Besides, deliberative strategies ($CC$, $CD$) deduct a reasoning cost $\gamma$, capturing the 
cognitive effort of conditional or strategic thinking, whereas intuitive strategies ($UC$, $UD$) incur no such cost. \tb{We constrain $0 \leq \gamma \leq 1$ to ensure the cost remains below the  benefit of mutual cooperation. The model approximates a classic Prisoner's Dilemma in two limiting conditions. When the reasoning cost $\gamma=0$, all strategies are costless, allowing deliberative types to exploit their conditional logic without penalty. When $\gamma$ is sufficiently high, deliberative strategies are suppressed due to their cost, and the dynamics reduce to a Prisoner's Dilemma like game between unconditional cooperators and defectors. }

In summary, considering the cheap talk stage and the subsequent actions in PDG for our chosen strategies, the payoff matrix is:
\begin{equation}
    \begin{tabular}{c|cccc}
           & \textit{UC} &  \textit{CC} &  \textit{UD} & \textit{CD} \\
           \hline
         \textit{UC} & $1$ & $1$ &$-r$ &$-r$\\
         \textit{CC} & $1-\gamma$ & $1-\gamma$ &$0-\gamma$ &$-r-\gamma$\\
         \textit{UD} & $1+r$ & $0$ &$0$ &$0$ \\
         \textit{CD} & $1+r-\gamma$ & $1+r-\gamma$ &$0-\gamma$ &$0-\gamma$ 
    \end{tabular}
    \label{payoffMatrix}
\end{equation}
where row strategies denote the focal player’s strategy, and column strategies denote the co-player’s strategy. 

\subsection{Well-mixed finite population}
We  consider a well-mixed finite population consisting of $M$ players. Each player updates their strategy following the Moran process. At each time step, a randomly selected player updates its strategy by imitating the strategy of another randomly selected player. 
Suppose there are only two strategies, $A$ and $B$, in the population, which can be one of the four strategies $UC$, $CC$, $UD$, and $CD$.
Assuming there are $m$ players adopting strategy $A$ (i.e. $M-m$ players adopting $B$), then the average payoffs for $A$ and $B$ are, respectively 
\begin{equation}
\begin{split}
     & f_{A}=\frac{(m-1)\pi_{A,A}+(M-m)\pi_{A,B}}{M-1}, \\
     & f_{B}=\frac{m\pi_{B,A}+(M-m-1)\pi_{B,B}}{M-1},
\end{split}
\end{equation}
where $\pi_{A,B}$ denotes the payoff when the focal player adopting strategy $A$ interacts with a co-player adopting strategy $B$.
We assume the evolutionary dynamic is driven by the Fermi function, one of the typical social learning rules, where players tend to imitate the strategy of those with  a larger payoff~\cite{sigmund2010social}. In detail, a player with payoff $f_A$ imitates the strategy of the pairwise player with payoff $f_B$ with probability, $(1+e^{s (f_{A}-f_{B})})^{-1}$. Therein $s$ represents the selection intensity, determining how strongly players rely on the payoff difference when making their imitation decision. When $s\rightarrow \infty$, imitation is deterministic, while when $s=0$, it becomes the neutral drift where players randomly adopt an existing strategy.

Based on the above assumption, the probability of the number $m$ of strategy $A$ in the population increasing or decreasing by 1 is:
\begin{equation}
       T_{AB}^{\pm}=\frac{M-m}{M}\frac{m}{M}[1+e^{\mp s (f_{A}-f_{B})}]^{-1}.
\end{equation}
The fixation probability of a mutant with strategy $A$ in a resident population of $M-1$ $B$-players is \cite{traulsen2006stochastic}:
\begin{equation}
    \rho_{BA}=\frac{1}{1+\sum_{k=0}^{M-1}\prod_{m=1}^{k}\frac{T^-_{AB}}{T^+_{BA}}}.
\end{equation}
Assuming a small mutation limit, where  any mutant  either fixates or goes extinct before another mutation occurs \cite{nowak2004emergence,imhof2005evolutionary},   the fixation probabilities $\rho_{AB}$ define the transition probabilities of the Markov process between four different homogeneous states of the population. 
The transition matrix with $T_{AB,A\neq B}=\rho_{AB}/(q-1)$ and $T_{AB}=1-\sum_{B=1, B\neq A}^qT_{AB}$, where $q$ is the number of strategies. The normalised eigenvector associated with the eigenvalue 1 of the transposed transition matrix provides the stationary distribution described above, describing the relative time the population spends adopting each of the strategies.

\subsection{Lattice finite population}
Different from the well-mixed population where players interact globally, lattice populations involve local interactions between neighbours.
We model a two-dimensional regular lattice with periodic boundary conditions, where each node is occupied by one player, and each player can only interact with its von Neumann neighbourhood along edges. 
The simulation proceeds as follows: players are initially assigned one of the four strategies, $UC$, $CC$, $UD$, or $CD$, with equal probability. Each player accumulates payoffs by interacting with its four neighbours using the payoff matrix in Eq. (\ref{payoffMatrix}). The strategy update process follows the asynchronous Monte Carlo algorithm: in each step, a randomly selected player $i$ either mutates to a random strategy with probability $\mu$, or imitates a randomly chosen neighbour $j$'s strategy with probability $1-\mu$. The imitation probability follows the Fermi function:
\begin{equation}
    F(j\rightarrow i)=\frac{1}{1+e^{(\phi_i-\phi_j)k^{-1}}},
    \label{eq:fermi}
\end{equation}
where $\phi_i$ and $\phi_j$ are the payoffs of player $i$ and $j$, respectively, and $k^{-1}$ corresponds to the selection strength, determining how sensitive the strategy update is to payoff differences. 
To align with strong selection scenarios in well-mixed populations, we fix $k^{-1}=10$. A Monte Carlo step consists of $L^2$ updates, where $L^2$ is the size of the population. This ensures each player updates their strategy on average once. Simulations run for $3\times10^4$ steps, with results averaged over the last 3,000 steps after confirming evolutionary equilibrium.  The population size is $L^2=200^2$ unless specified.
A mutation $\mu=10^{-5}$ is incorporated into the strategy-updating to prevent the finite-size effect ~\cite{shen2025mutation}.

\begin{figure*}[tb]
    \centering
    \includegraphics[width=1\textwidth]{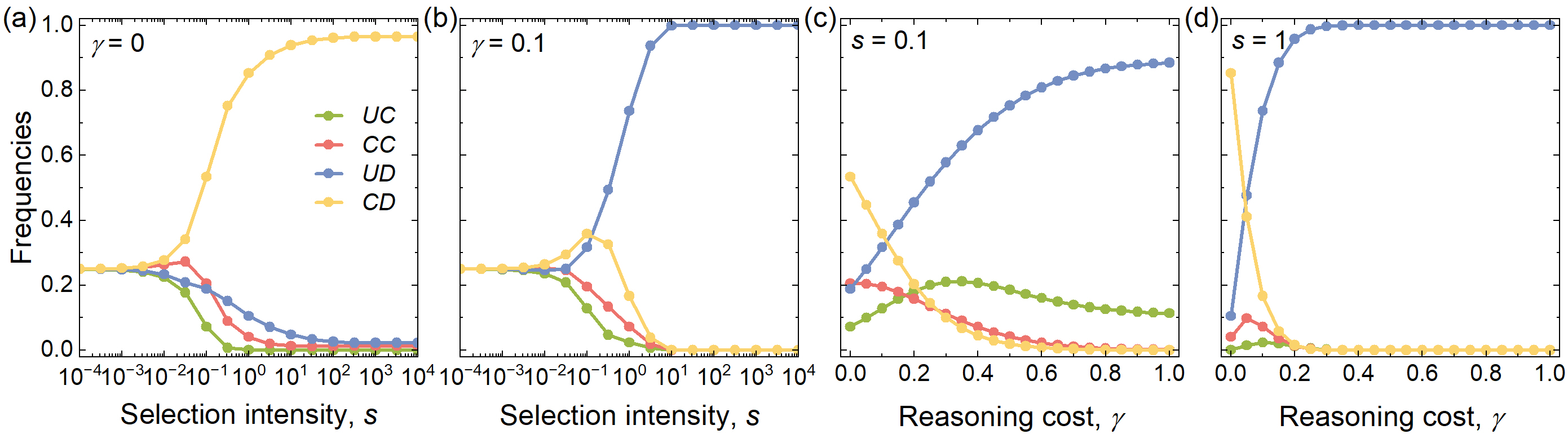}
    \caption{
    \textbf{Cheap talk cannot sustain cooperation in well-mixed finite populations unless the selection intensity is weak. }
    Panels (a) and (b) show the stationary distributions of each strategy against the selection intensity $s$, with no reasoning cost ($\gamma=0$) and with small reasoning cost ($\gamma=0.1$), respectively. Panels (c) and (d) show the stationary distributions of each strategy against the reasoning cost when selection intensity is weak ($s=0.1$) and strong ($s=1$), respectively.  
    The parameter is set as $r=0.2$.}
    \label{fig:finite}
\end{figure*}

\section{Results}
\subsection{Well-mixed  population}
In the one-shot, anonymous game in a well-mixed finite population, in which reciprocity mechanisms involving network reciprocity are absent ~\cite{rand2013human}, cheap talk has limited effectiveness in sustaining cooperation ($UC$ and $CC$), primarily succeeding only under a weak selection intensity.
\tb{
When the reasoning cost is absent ($\gamma=0$, Figure \ref{fig:finite}(a)), at a weak selection intensity (around $s \leq 10^{-2}$), all four strategies---$UC$, $CC$, $UD$, and $CD$---co-exist with frequencies around 0.25. However, as the selection intensity increases, $CD$ begins to dominate, while other strategies decline.
When a small reasoning cost is considered ($\gamma=0.1$, Figure \ref{fig:finite} (b)), a similar co-existence can be found under weak selection intensities, but as the selection intensifies, $CD$ first increases to a peak and then diminishes to extinction, while $UD$ steadily increases to dominance.}
Furthermore, for a weak selection intensity ($s=0.1$, Figure \ref{fig:finite}(c)), $UC$ persists regardless of the reasoning cost, and $CC$ remains until reasoning cost becomes large (around $\gamma \geq 0.8$). In contrast, under strong selection intensity ($s=1$, Figure \ref{fig:finite}(d)), both $UC$ and $CC$ briefly rise to a low peak at a small reasoning cost before declining to zero (around $\gamma=0.2$). 

These results show the minimal benefits of cheap talk for the cooperation dominance in well-mixed populations.  
Although $UC$ and $CC$ can survive under weak selection intensities, strong selection undermines the effectiveness of cheap talk.
This aligns with laboratory evidence showing that although cheap talk shortly boosts cooperation, it cannot sustain cooperative behaviour in the long-term within the one-shot Prisoner’s Dilemma game ~\cite{duffy2002actions}.
In essence, cheap talk fails to alter the evolutionary equilibrium in such homogeneous settings.
Notably, in contrast to well-mixed populations, network reciprocity allows cooperation to survive by forming clusters to avoid the invasion of defection in structured populations~\cite{nowak1992evolutionary}.
Such differences further motivate  our exploration of the impact of cheap talk on structured populations, that is, the interplay of cheap talk and network reciprocity.

\begin{figure*}[tb]
    \centering
    \includegraphics[width=0.5\linewidth]{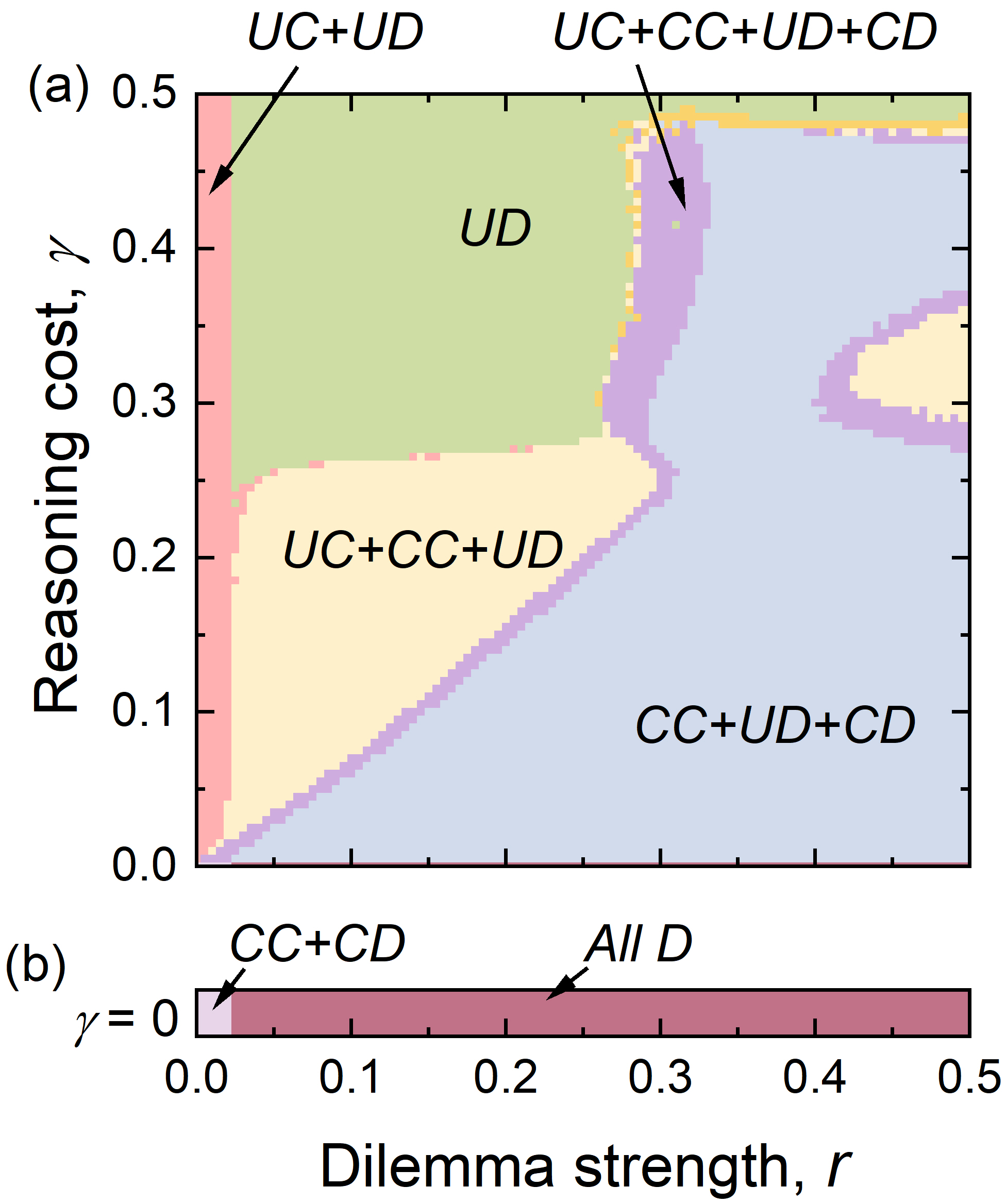}
    \caption{
    \textbf{Cheap talk sustains cooperation when it adheres to network reciprocity.}
    Panel (a) shows the full $r-\gamma$ phase diagram obtained by Monte Carlo simulations on the square lattice network. Five main co-existences of strategies, including $UD$, $UC+UD$, $UC+CC+UD$, $CC+UD+CD$, and $UC+CC+UD+CD$, are coloured as green, red, yellow, blue, and purple, respectively.
    Panel (b) shows the zoomed-in results when reasoning cost $\gamma=0$. Co-existence of $CC+CD$ is coloured as light purple, and $All D$ where only $UD$ or $CD$ exists or co-exists, is coloured in dark red.
    }
    \label{fig:phaseDiagram}
\end{figure*}

\begin{figure*}[htb]
    \centering
{\includegraphics[width=1\textwidth]{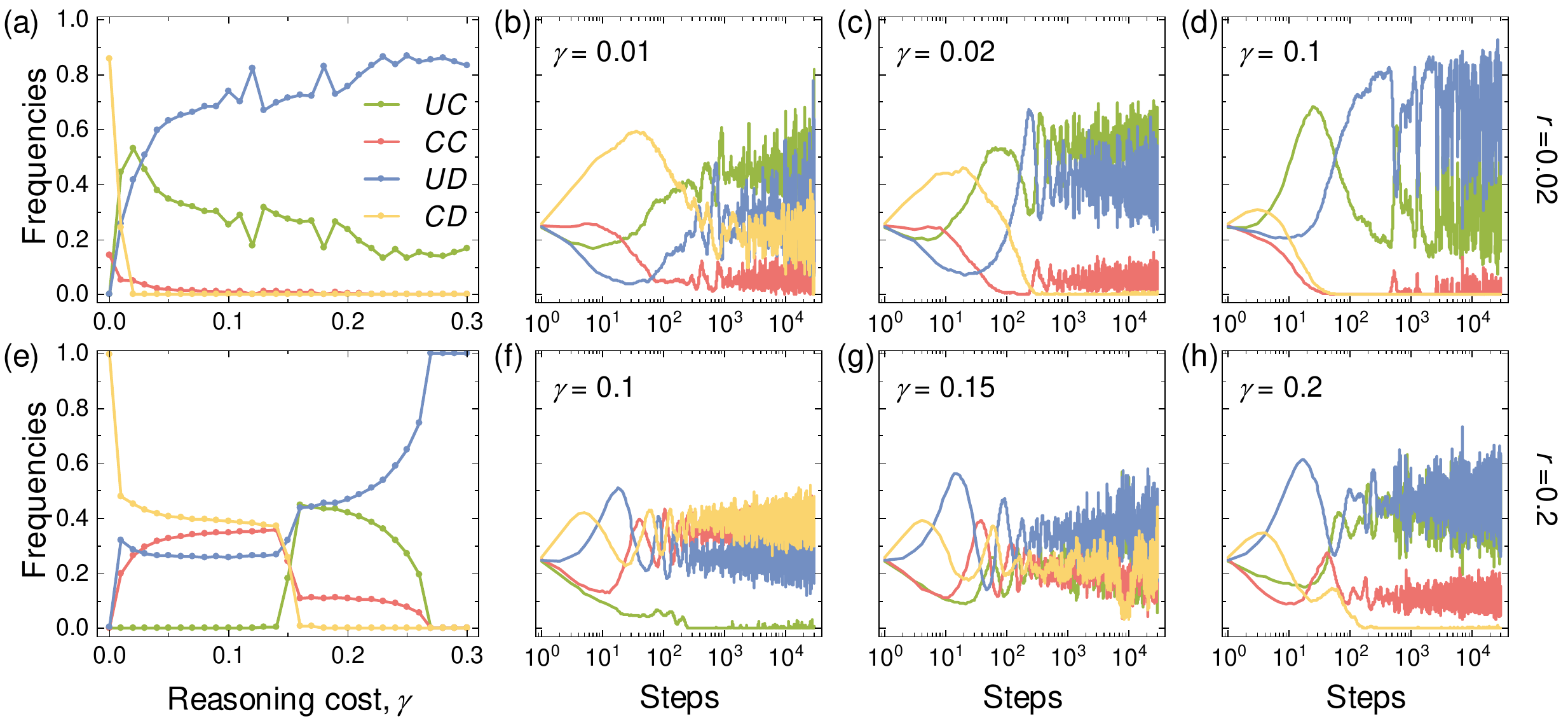}} 
    \caption{
    \textbf{
   Optimal ranges of the reasoning cost bring about and sustain cooperation by various co-existence states.}
    Shown are the frequencies of each strategy as a function of reasoning cost $\gamma$ in the first column, and the frequencies of each strategy over time in the other columns. 
    Parameters are set $r=0.02$ in the top row, $r=0.2$ in the bottom row, and (b) $\gamma=0.01$, (c) $\gamma=0.02$, (d) $\gamma=0.1$, (f) $\gamma=0.1$, (g) $\gamma=0.15$, (h) $\gamma=0.2$. }
    \label{fig:evolution}
\end{figure*}

\begin{figure*}[htb]
    \centering
    \subfigure
    {\includegraphics[width=0.9\textwidth]{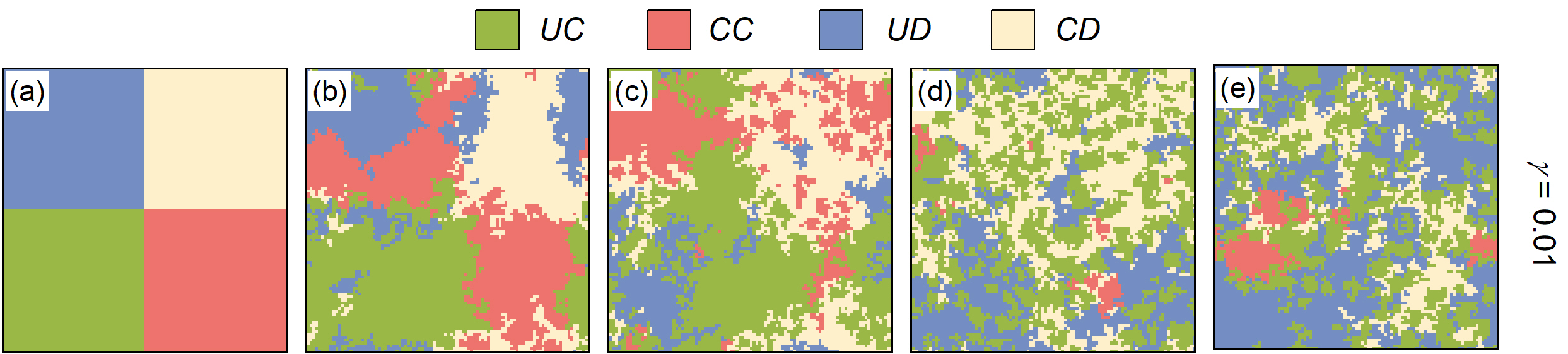}} \subfigure{\includegraphics[width=0.9\textwidth]{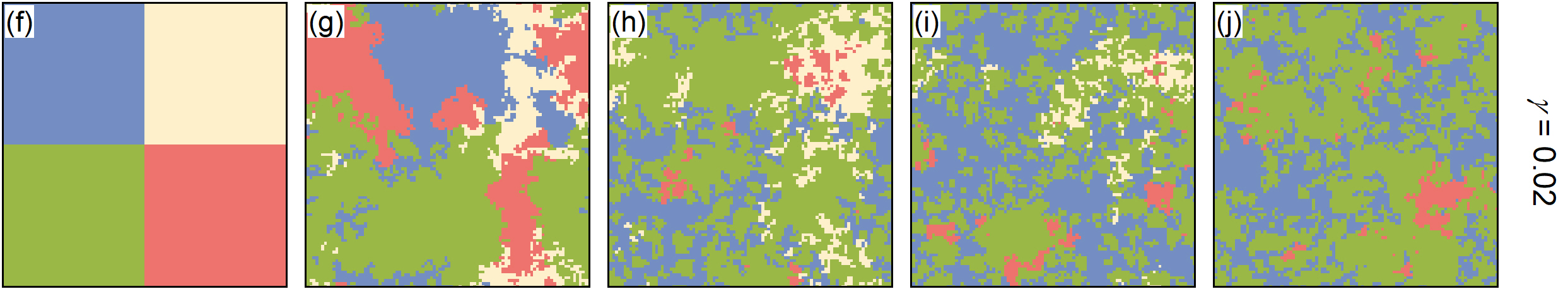}} 
\subfigure{\includegraphics[width=0.9\textwidth]{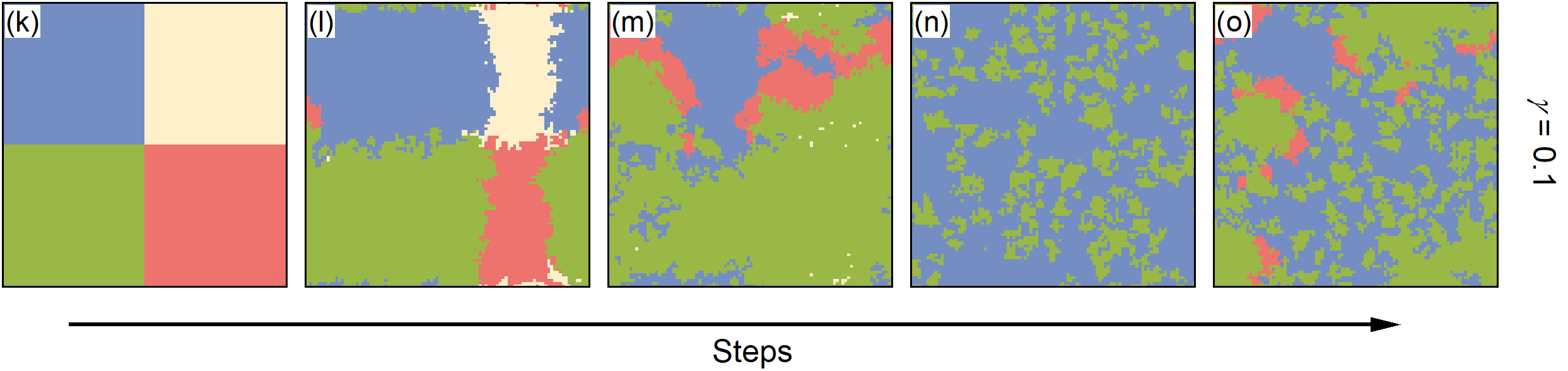}} 
    \caption{
    \textbf{
    Conditional cooperation ($CC$) acts as the catalyst for unconditional cooperation ($UC$) at weak.}
    Shown are evolutionary snapshots at specific time steps (columns) and for different reasoning costs (rows).
    \tb{
    With the dilemma strength $r=0.02$, the top row shows the evolution for a reasoning cost of  $\gamma=0.01$ at time steps 0, 70, 150, 650, and 800. In the middle row, the cost is  $\gamma=0.02$ with snapshots at time steps 0, 80, 200, 300, and 700. The bottom row corresponds to a cost of  $\gamma=0.1$ and is shown at time steps 0, 30, 80, 300, and 1350.}
    }
    \label{fig:snapshot1}
\end{figure*}

\begin{figure*}[htb]
    \centering
    \subfigure
    {\includegraphics[width=0.9\textwidth]{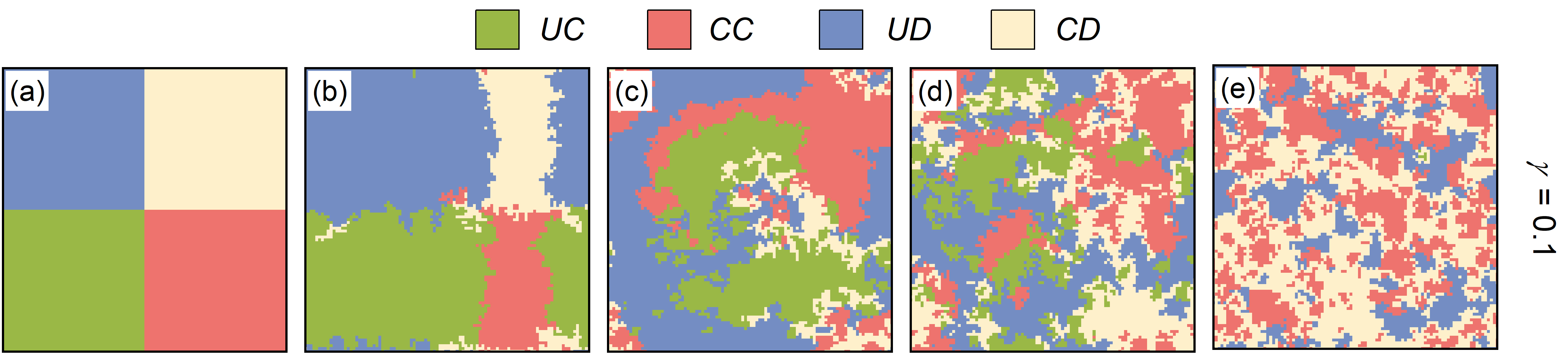}} \subfigure{\includegraphics[width=0.9\textwidth]{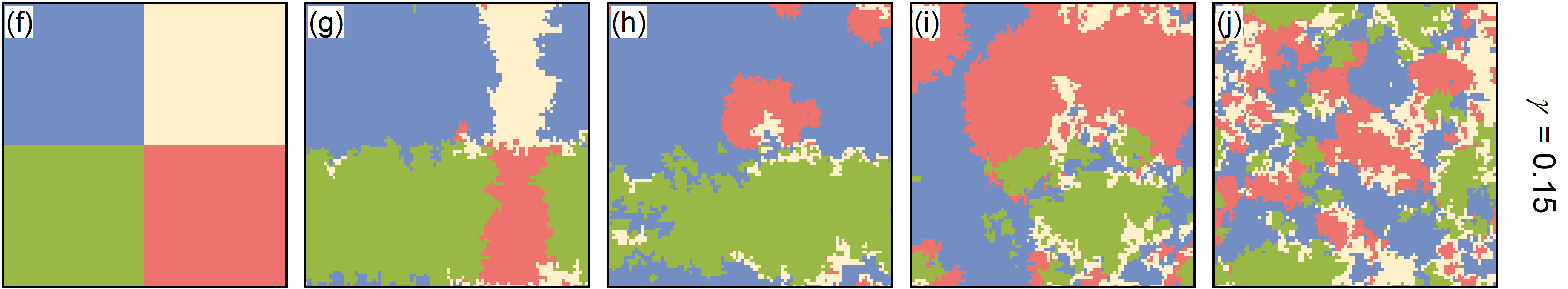}} 
\subfigure{\includegraphics[width=0.9\textwidth]{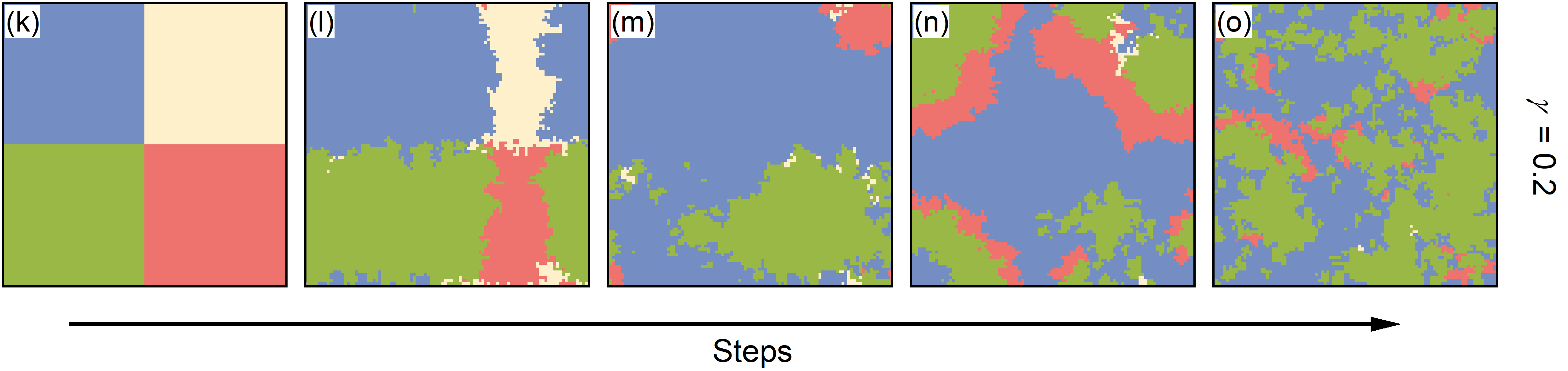}} 
    \caption{
    \textbf{
    At strong dilemma strength, $CC$ remains the catalyst of $UC$, though challenged by strategic defection $CD$. }
    \tb{
    Shown are evolutionary snapshots at specific time steps (columns) and for different reasoning costs (rows).
    With the dilemma strength $r=0.2$, the top row shows the evolution for a reasoning cost of  $\gamma=0.1$ at time steps 0, 30, 90, 200, and 1000. In the middle row, the cost is  $\gamma=0.15$ with snapshots at time steps 0, 30, 60, 100, and 1000. The bottom row corresponds to a cost of  $\gamma=0.2$ and is shown at time steps 0, 30, 90, 150, and 1000.}
    }
    \label{fig:snapshot2}
\end{figure*}

\subsection{Structured population}

Together with network reciprocity, cheap talk sustains cooperation via various complex co-existing states, mainly including $UC+UD$, $UC+CC+UD$, $CC+UD+CD$, and $UC+CC+UD+CD$. 
\tb{
When the reasoning cost is negligible ($\gamma=0$, Figure \ref{fig:phaseDiagram}(b)), $CC$ co-exists with $CD$ when $r\leq0.02$. Beyond this, defection tends to dominate.
Conversely, the increasing reasoning cost extends the parameter space for cooperation strategies, $UC$ and $CC$ (Figure \ref{fig:phaseDiagram}(a)).}
In detail, when the dilemma strength is small (around $r\leq0.02$), four strategies co-exist before $CD$ and then $CC$ are driven to extinction as the reasoning cost increases.
When the dilemma strength is moderate (around $r\leq0.25$), a transition from the co-existence of $CC$, $UD$ and $CD$, to the emergence of $UC$, then the extinction of $CD$, and eventual $UD$ dominance with a rising reasoning cost. 
When the dilemma strength becomes larger, cheap talk allows $CC$ to persist with $UD$ and $CD$ (and sometimes $UC$) across a broad range of parameter values until a high reasoning cost drives $UD$ to dominance.
These results demonstrate that the interplay of cheap talk and network reciprocity sustains cooperation even under strong dilemma strengths and high reasoning costs, far beyond what is possible under each mechanism (cheap talk or network reciprocity) individually.

Notably, minor deviations near state boundaries due to the finite-size effect, large-population simulations in Figure \ref{fig:mutation} confirm these points belong to the main co-existence states rather than uncovered regimes. We incorporated small mutations to address finite-size effects during simulations ~\cite{shen2025mutation}, though this approach is insufficient to fully resolve discrepancies under our current context.
We retain results from the current population size, as it is computationally feasible and sufficiently demonstrates how cheap talk, combined with network reciprocity, robustly sustains cooperation across diverse parameter regimes.

To better understand how cheap talk influences the co-existence states, we examine the frequencies of each strategy as a function of reasoning cost and present key evolutionary trajectories in Figure \ref{fig:evolution}. Our results reveal an optimal range of reasoning costs where cooperation co-exists with other strategies. Furthermore, the co-existence of cooperation depends on reasoning costs as well as the dilemma strength.
At weak dilemma strength, cheap talk allows $UC$ to co-exist with others and even dominate in the population. On the other hand, at strong dilemma strength, the fate of cooperation relies on the reasoning cost: small reasoning costs allow $CC$ to persist as an equilibrium, while large reasoning costs enable both $UC$ and $CC$ to persist in equilibrium with defection.

At a weak dilemma strength, cheap talk enables $UC$ to thrive across varying reasoning costs, but increasing costs progressively hinder $CC$, eroding the effectiveness of cheap talk and reverting dynamics to those sustained by network reciprocity alone.
In Figure \ref{fig:evolution}(a),  \tb{when the reasoning cost is negligible, $CC$ co-exists with $CD$ at a low frequency. With the increase in reasoning cost, $UC$ and $UD$ become prevail, but $CC$ and $CD$ are gradually eliminated (at around $\gamma=0.02$ and $\gamma=0.22$, respectively); $UC$ dominates at relatively low reasoning costs (around $\gamma \leq 0.02$) before declining, while $UD$ gradually increases. }
At a high reasoning cost (around $\gamma=0.3$), \tb{only $UC$ and $UD$ survive, echoing the coexistence of $CC$ and $CD$ in the absence of reasoning costs ($\gamma=0$).}
In more specific scenarios, when four strategies co-exist ($\gamma=0.01$, Figure \ref{fig:evolution}(b)), $CC$ decreases steadily over time. Meanwhile, $UC$, $UD$ and $CD$ show complex fluctuations, with $UD$ decreasing initially and then increasing slightly, and $UC$ and $CD$ showing the opposite trend. Eventually, $UC$ emerges as the dominant strategy, relegating the others to lower frequencies.
When $CD$ goes extinct and the remaining three strategies co-exist ($\gamma=0.02$, Figure \ref{fig:evolution}(c)), the evolutionary pattern mirrors that of $\gamma=0.01$, with $CD$ being eliminated after the fluctuation, and $UC$ and $UD$ co-exist with $CC$ at higher frequencies by the end.
At $\gamma=0.1$ (Figure \ref{fig:evolution}(d)), in the scenario where $CD$ is extinct and the other three strategies co-exist, a distinct pattern unfolds: $CC$ nearly extincts yet experiences frequent upward surges, while $UD$ persists at a relatively high frequency compared to the others.

At strong dilemma strengths, high reasoning costs challenge the survival of $CC$ and $CD$ while paradoxically enabling $UC$---which typically struggles to emerge under such conditions---to persist with cheap talk.
As depicted in Figure \ref{fig:evolution}(e), \tb{when the reasoning cost is negligible, $CD$ dominates in the population. A small reasoning cost allows $CC$ and $UD$ to emerge and co-exist with $CD$ (around $\gamma \leq 0.14$).
With a moderate reasoning cost (around $0.14<\gamma<0.16$), $UC$ appears as $CD$ and $CC$ frequencies decrease, leading to the co-existence of four strategies. Subsequently, $CD$ vanishes, resulting in the coexistence of the remaining three strategies (around $0.16\le \gamma<0.26$). At a high reasoning cost (around $0.27\le \gamma$), $UD$ eventually dominates the population, which is similar to the prevalence of defection when the reasoning cost is negligible.}
In detail, when $CC$, $UD$, and $CD$ co-exist ($\gamma=0.1$, Figure \ref{fig:evolution}(f)), $UC$ is gradually eliminated as the evolution process unfolds, while $CC$, $UD$, and $CD$ experience fluctuations and eventually co-exist at similar frequencies. 
When four strategies co-exist ($\gamma=0.15$, Figure \ref{fig:evolution}(g)), $UC$ survives after fluctuations, similar to the other three strategies. 
When four strategies co-exist ($\gamma=0.2$, Figure \ref{fig:evolution}(h)), $CD$ vanishes after an upward fluctuation, while the remaining strategies co-exist. 

These results confirm that reasoning cost is central to determining the co-existence of cooperation strategies with other strategies, as well as the frequency dynamics in the population. To reveal the inherent dynamics and the competitive interplay among strategies, we record snapshots of strategy distribution during the evolutionary processes \footnote{see detailed dynamics at: \url{https://osf.io/zpdc4/?view_only=938bc47b057c4ddeb3d2e9440c4a1d65}} and visualise typical steps in Figure \ref{fig:snapshot1} and Figure \ref{fig:snapshot2}. Our results further uncover the pivotal role of $CC$ in the prevalence of $UC$, a universal catalyst that persists even when challenged by $CD$ under strong dilemma strength.
Note that we use the prepared initial distribution to improve the visibility of strategic relationships; importantly, the evolutionary outcomes show no qualitative differences from those arising from random initial distributions.

At a weak dilemma strength ($r=0.02$, Figure \ref{fig:snapshot1}), $UC$ persists in the population through the cyclic dominance among $UC$, $CC$, and $UD$, where $UC$ replaces $CC$, $CC$ replaces $UD$, and $UD$ replaces $UC$, driven by $CC$, even at a low frequency. 
At a small reasoning cost ($\gamma=0.01$, the top row in Figure \ref{fig:snapshot1}), cyclic dominance sustains $UC$ and $CC$, while $CD$ survives by exploiting cooperation. 
As the reasoning cost increases ($\gamma=0.02$, the middle row in Figure \ref{fig:snapshot1}), $CD$ is eliminated, but $CC$ survives through the cyclic dominance. 
At a large reasoning cost ($\gamma=0.1$, the bottom row in Figure \ref{fig:snapshot1}), $CC$ approaches  extinction, and $UD$ exists in small clusters and at a low frequency. However, the small mutation preserves enough $CC$ to sustain cyclic dominance, benefiting the prevalence of $UC$.
To conclude, these results highlight the pivotal role of $CC$  in fostering the prevalence of $UC$ by cyclic dominance. Importantly, the structured population setting contributes to the formation of  $CC$ clusters, avoiding the vanishing alongside $CD$ due to the reasoning cost, highlighting the synergistic role of cheap talk and network reciprocity in sustaining cooperation.

At a strong dilemma strength ($r=0.2$, Figure \ref{fig:snapshot2}), $CC$ navigates dual cyclic dominance: one facilitates the survival of $UC$, while another among $CC$, $UD$, and $CD$ might pose challenges---where $CD$ replaces $CC$, $CC$ replaces $UD$, and $UD$ replaces $UC$.
At a low reasoning cost (the top row in Figure \ref{fig:snapshot2},  $\gamma=0.1$), $UC$ is eliminated by $UD$ and $CD$, but $CC$ is competitive against defection by cyclic dominance with $UD$ and $CD$.
As reasoning cost increases (the middle row in Figure \ref{fig:snapshot2},  $\gamma=0.15$), $CD$ becomes less competitive compared to $UD$, allowing $UC$ to re-emerge via a second cyclic dominance with $CC$ and $UD$; in this case, two cyclic dynamics co-exist in the population. 
At a high reasoning cost (the bottom row in Figure \ref{fig:snapshot2},  $\gamma=0.2$), $CD$ is rapidly replaced by $UD$, leaving the single cyclic dominance among $UC$, $CC$, and $UD$.
Together with previous results, cheap talk sustains cooperation at proper reasoning cost even under strong dilemma strength, leveraging $CC$ as a mediator in cyclic dominance (among $UC$, $CC$, and $UD$), with network reciprocity serving as the indispensable scaffold for the  evolutionary stability of $CC$. These findings underscore how cheap talk and network reciprocity jointly shape the fate of cooperation, revealing the conditional yet profound role of communication in complex systems.

\section{Discussion}
We have investigated the evolutionary role of non-binding communication in promoting cooperation by developing a game-theoretic model inspired by dual-process cognitive theory. Although behavioural experiments showed that cheap talk can increase cooperation, its effectiveness is puzzling from an evolutionary perspective, since it does not alter payoffs and should not affect strategy selection. Consistent with this, we found that cheap talk cannot sustain cooperation in well-mixed populations under one-shot, anonymous interactions. However, when embedded in structured populations, it becomes effective: cooperation emerges through cyclic dominance among multiple strategies, with conditional cooperators acting as catalysts that protect unconditional cooperators. This mechanism is most effective when social reasoning costs are moderate, with cooperation surviving in the form of conditional or unconditional cooperators within cyclic dominance patterns sensitive to these costs. Outside this range, cooperation persists via baseline network reciprocity or collapses. Our findings offer a new evolutionary explanation for why, and under what conditions, non-binding communication can shape cooperation.

Beyond uncovering a specific mechanism, our study offers a general evolutionary framework for understanding how non-binding communication influences cooperation. While traditional models have focused on enforceable communication---where signals are tied to material costs or incentives---our framework shows that costless signals can still shape cooperative outcomes when combined with cognitive heterogeneity and structured interactions. It reproduces well-established results in well-mixed, one-shot settings ~\cite{duffy2002actions} and reveals that, in structured populations, network reciprocity enables cheap talk to sustain cooperation through complex patterns of cyclic dominance. These findings underscore the framework’s utility in capturing the evolutionary consequences of non-binding communication. Moreover, it can be extended to settings where cheap talk has demonstrated empirical success but lacks theoretical grounding, including repeated games, public goods dilemmas, and beyond. It also provides a baseline for comparing the evolutionary impact of binding versus non-binding communication and for extending the model to human–AI hybrid systems by incorporating preprogrammed agents~\cite{dafoe2020open}.

Our model adopts a deliberately minimal design to isolate core mechanisms. To reduce complexity, we focused on four representative strategies ($UC$, $UD$, $CC$, $CD$), although the two-stage Donation game permits eight in total. Including additional strategies would increase dynamical complexity without altering the main result, as the catalytic role of $CC$ remains robust due to preserved cyclic dominance with $UC$ and $UD$. A larger strategy space may introduce more intricate power relations among strategies and lead to richer dynamics, warranting further exploration in both structured and well-mixed populations. We also assume equal cognitive costs for deliberative strategies and none for intuitive ones, consistent with dual-process theory, which distinguishes effortful, context-sensitive reasoning from automatic responses~\cite{bear2016intuition}. Varying these costs could shift the relative fitness of strategies and potentially lead to alternative dynamics, which presents a natural direction for further theoretical investigation.

Finally, our results are based on evolutionary simulations conducted on square lattice networks, which serve as a standard but stylised representation of local interactions~\cite{nowak2006five}. This structure supports the formation of cooperative clusters that protect individuals from exploitation, illustrating a key mechanism of network reciprocity. Although simplified, it captures essential differences between local and global interaction and shows how non-binding communication can gain evolutionary traction through spatial reinforcement. Future work could extend this framework by examining weaker selection regimes, where individuals are less responsive to payoff differences, or by incorporating noise in signalling and action execution ~\cite{han2022institutional}. Additional extensions might consider more realistic features such as scale-free networks, dynamic topologies, or higher-order interactions to better reflect the complexity of real-world social systems.

\section*{Acknowledgments}
The idea was initiated and formalised during Dr. Shen’s research stay at Teesside University. We acknowledge the support provided by EPSRC (grant EP/Y00857X/1) to Z.S. and T.A.H., and JSPS KAKENHI (Grant no. JP 23H03499) to C.\,S..

\section*{Author contributions}
C.S. conceived and designed the study; Z.S. and C.S. performed research; all authors analyzed results and wrote the manuscript.

\section*{Competing interest} Authors declare that they have no conflict of interest.

\section*{Data availability}
No datasets were generated or analysed during the current study. The code to support the findings of this
study is available at \url{https://osf.io/zpdc4/?view_only=938bc47b057c4ddeb3d2e9440c4a1d65}. 

\clearpage


\section*{Appendix}
\renewcommand{\thefigure}{A\arabic{figure}}
\renewcommand{\thetable}{A\arabic{table}}
\setcounter{figure}{0} 
\setcounter{table}{0} 

\begin{table*}[htb]
    \centering
    \caption{Eight strategies in the two-stage game with cheap talk.}
    \begin{tabular}{c|ccc}
    \toprule
     Strategy   & \makecell{Signal  \\cooperative intention?} & \makecell{Cooperate  \\if the co-player signals?} & \makecell{Cooperate \\ if the co-player remains silent?}\\
     \hline
    $SCC$     & yes & yes & yes \\
    $SCD$     & yes & yes & no \\
    $SDC$     & yes & no & yes \\
    $SDD$     & yes & no & no \\
    $NCC$     & no & yes & yes \\
    $NCD$     & no & yes & no \\
    $NDC$     & no & no & yes \\
    $NDD$     & no & no & no \\
    \bottomrule
    \end{tabular}
    \label{tab:8strategy}
\end{table*}

\begin{figure*}[h] 
    \centering
    \includegraphics[width=0.8\textwidth]{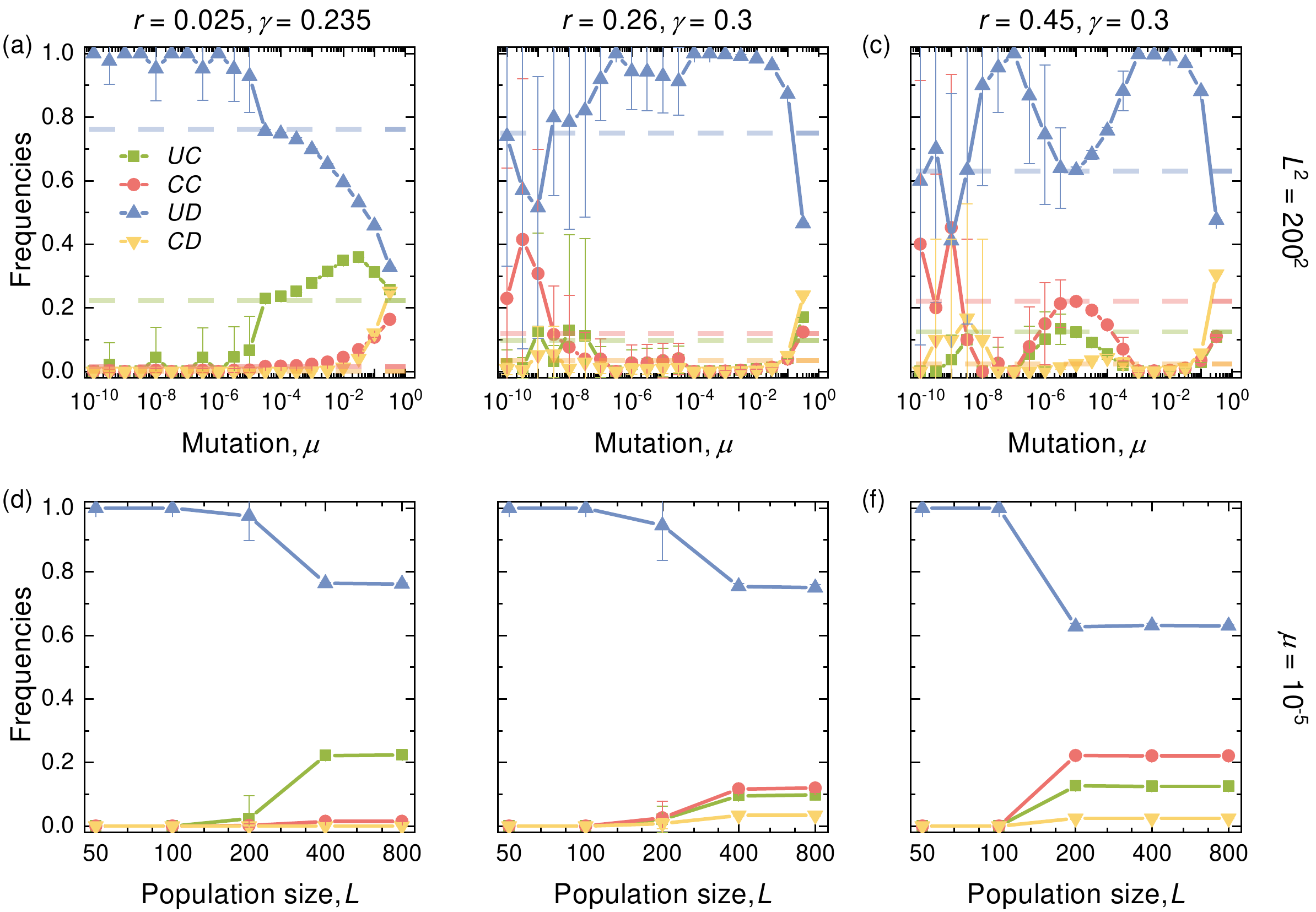} 
    \caption{ 
    \textbf{Finite-size deviations confirm alignment with the main co-existence states.}
    Shown are the frequency of each strategy as a function of mutation in the top row, and as a function of population size in the bottom row. Error bars indicate standard deviations across 20 independent runs.
    The horizontal dashed line represents reference outcomes from a large population ($L^2=800^2$).
    In the top row, population size $L^2=200^2$, in the bottom row, mutation $\mu=10^{-5}$. Parameters are set as $r=0.025$, $\gamma=0.235$ in the first column, $r=0.26$, $\gamma=0.3$ in the second column, and $r=0.45$, $\gamma=0.3$ in the last column.}
    \label{fig:mutation}
\end{figure*}
\clearpage
\bibliographystyle{unsrt}
\bibliography{mybib}

\begin{thebibliography}{10}

\bibitem{victor2022determining}
David~G Victor, Marcel Lumkowsky, and Astrid Dannenberg.
\newblock Determining the credibility of commitments in international climate policy.
\newblock {\em Nature Climate Change}, 12(9):793--800, 2022.

\bibitem{bavel2020using}
Jay J~Van Bavel, Katherine Baicker, Paulo~S Boggio, Valerio Capraro, Aleksandra Cichocka, Mina Cikara, Molly~J Crockett, Alia~J Crum, Karen~M Douglas, James~N Druckman, et~al.
\newblock Using social and behavioural science to support covid-19 pandemic response.
\newblock {\em Nature human behaviour}, 4(5):460--471, 2020.

\bibitem{smith2010communication}
Eric~Alden Smith.
\newblock Communication and collective action: language and the evolution of human cooperation.
\newblock {\em Evolution and human behavior}, 31(4):231--245, 2010.

\bibitem{dietz2003struggle}
Thomas Dietz, Elinor Ostrom, and Paul~C Stern.
\newblock The struggle to govern the commons.
\newblock {\em science}, 302(5652):1907--1912, 2003.

\bibitem{boyd2009culture}
Robert Boyd and Peter~J Richerson.
\newblock Culture and the evolution of human cooperation.
\newblock {\em Philosophical Transactions of the Royal Society B: Biological Sciences}, 364(1533):3281--3288, 2009.

\bibitem{rand2013human}
David~G Rand and Martin~A Nowak.
\newblock Human cooperation.
\newblock {\em Trends in cognitive sciences}, 17(8):413--425, 2013.

\bibitem{perc2017statistical}
Matja{\v{z}} Perc, Jillian~J Jordan, David~G Rand, Zhen Wang, Stefano Boccaletti, and Attila Szolnoki.
\newblock Statistical physics of human cooperation.
\newblock {\em Physics Reports}, 687:1--51, 2017.

\bibitem{hamilton1964genetical}
William~D Hamilton.
\newblock The genetical evolution of social behaviour. ii.
\newblock {\em Journal of theoretical biology}, 7(1):17--52, 1964.

\bibitem{crawford1998survey}
Vincent Crawford.
\newblock A survey of experiments on communication via cheap talk.
\newblock {\em Journal of Economic theory}, 78(2):286--298, 1998.

\bibitem{andersson2012credible}
Ola Andersson and Erik Wengstr{\"o}m.
\newblock Credible communication and cooperation: experimental evidence from multi-stage games.
\newblock {\em Journal of Economic Behavior \& Organization}, 81(1):207--219, 2012.

\bibitem{feldhaus2016more}
Christoph Feldhaus and Julia Stauf.
\newblock More than words: the effects of cheap talk in a volunteer’s dilemma.
\newblock {\em Experimental Economics}, 19:342--359, 2016.

\bibitem{duffy2002actions}
John Duffy and Nick Feltovich.
\newblock Do actions speak louder than words? an experimental comparison of observation and cheap talk.
\newblock {\em Games and Economic Behavior}, 39(1):1--27, 2002.

\bibitem{erbaugh2024communication}
James~T Erbaugh, Charlotte~H Chang, Yuta~J Masuda, and Jesse Ribot.
\newblock Communication and deliberation for environmental governance.
\newblock {\em Annual Review of Environment and Resources}, 49, 2024.

\bibitem{charness2006promises}
Gary Charness and Martin Dufwenberg.
\newblock Promises and partnership.
\newblock {\em Econometrica}, 74(6):1579--1601, 2006.

\bibitem{blume2007effects}
Andreas Blume and Andreas Ortmann.
\newblock The effects of costless pre-play communication: Experimental evidence from games with pareto-ranked equilibria.
\newblock {\em Journal of Economic theory}, 132(1):274--290, 2007.

\bibitem{tingley2011can}
Dustin~H Tingley and Barbara~F Walter.
\newblock Can cheap talk deter? an experimental analysis.
\newblock {\em Journal of Conflict Resolution}, 55(6):996--1020, 2011.

\bibitem{sutter2009communication}
Matthias Sutter and Christina Strassmair.
\newblock Communication, cooperation and collusion in team tournaments—an experimental study.
\newblock {\em Games and Economic Behavior}, 66(1):506--525, 2009.

\bibitem{balliet2010communication}
Daniel Balliet.
\newblock Communication and cooperation in social dilemmas: A meta-analytic review.
\newblock {\em Journal of Conflict Resolution}, 54(1):39--57, 2010.

\bibitem{sally1995conversation}
David Sally.
\newblock Conversation and cooperation in social dilemmas: A meta-analysis of experiments from 1958 to 1992.
\newblock {\em Rationality and society}, 7(1):58--92, 1995.

\bibitem{arechar2017m}
Antonio~A Arechar, Anna Dreber, Drew Fudenberg, and David~G Rand.
\newblock “i'm just a soul whose intentions are good”: the role of communication in noisy repeated games.
\newblock {\em Games and Economic Behavior}, 104:726--743, 2017.

\bibitem{dvorak2024negotiating}
Fabian Dvorak and Sebastian Fehrler.
\newblock Negotiating cooperation under uncertainty: Communication in noisy, indefinitely repeated interactions.
\newblock {\em American Economic Journal: Microeconomics}, 16(3):232--258, 2024.

\bibitem{ellingsen2010does}
Tore Ellingsen and Robert {\"O}stling.
\newblock When does communication improve coordination?
\newblock {\em American Economic Review}, 100(4):1695--1724, 2010.

\bibitem{bahel2022communication}
Eric Bahel, Sheryl Ball, and Sudipta Sarangi.
\newblock Communication and cooperation in prisoner's dilemma games.
\newblock {\em Games and Economic Behavior}, 133:126--137, 2022.

\bibitem{bernheim1987coalition}
B~Douglas Bernheim, Bezalel Peleg, and Michael~D Whinston.
\newblock Coalition-proof nash equilibria i. concepts.
\newblock {\em Journal of economic theory}, 42(1):1--12, 1987.

\bibitem{axelrod1981evolution}
Robert Axelrod and William~D Hamilton.
\newblock The evolution of cooperation.
\newblock {\em science}, 211(4489):1390--1396, 1981.

\bibitem{nowak2006evolutionary}
Martin~A Nowak.
\newblock {\em Evolutionary dynamics: exploring the equations of life}.
\newblock Harvard university press, 2006.

\bibitem{gintis2001costly}
Herbert Gintis, Eric~Alden Smith, and Samuel Bowles.
\newblock Costly signaling and cooperation.
\newblock {\em Journal of theoretical biology}, 213(1):103--119, 2001.

\bibitem{lang2024role}
Martin Lang, Radim Chvaja, and Benjamin~G Purzycki.
\newblock The role of costly commitment signals in assorting cooperators during intergroup conflict.
\newblock {\em Evolution and human behavior}, 45(2):131--143, 2024.

\bibitem{han2016emergence}
The~Anh Han.
\newblock Emergence of social punishment and cooperation through prior commitments.
\newblock In {\em Proceedings of the thirtieth aaai conference on artificial intelligence}, pages 2494--2500, 2016.

\bibitem{han2013emergence}
The~Anh Han and The~Anh Han.
\newblock The emergence of commitments and cooperation.
\newblock {\em Intention Recognition, Commitment and Their Roles in the Evolution of Cooperation: From Artificial Intelligence Techniques to Evolutionary Game Theory Models}, pages 109--121, 2013.

\bibitem{salahshour2019evolution}
Mohammad Salahshour.
\newblock Evolution of costly signaling and partial cooperation.
\newblock {\em Scientific reports}, 9(1):8792, 2019.

\bibitem{duong2021cost}
Manh~Hong Duong and The~Anh Han.
\newblock Cost efficiency of institutional incentives for promoting cooperation in finite populations.
\newblock {\em Proceedings of the Royal Society A}, 477(2254):20210568, 2021.

\bibitem{camerer2004cognitive}
Colin~F Camerer, Teck-Hua Ho, and Juin-Kuan Chong.
\newblock A cognitive hierarchy model of games.
\newblock {\em The Quarterly Journal of Economics}, 119(3):861--898, 2004.

\bibitem{bear2016intuition}
Adam Bear and David~G Rand.
\newblock Intuition, deliberation, and the evolution of cooperation.
\newblock {\em Proceedings of the National Academy of Sciences}, 113(4):936--941, 2016.

\bibitem{rubinstein2007instinctive}
Ariel Rubinstein.
\newblock Instinctive and cognitive reasoning: A study of response times.
\newblock {\em The Economic Journal}, 117(523):1243--1259, 2007.

\bibitem{axelrod1988further}
Robert Axelrod and Douglas Dion.
\newblock The further evolution of cooperation.
\newblock {\em Science}, 242(4884):1385--1390, 1988.

\bibitem{wang2015universal}
Zhen Wang, Satoshi Kokubo, Marko Jusup, and Jun Tanimoto.
\newblock Universal scaling for the dilemma strength in evolutionary games.
\newblock {\em Physics of life reviews}, 14:1--30, 2015.

\bibitem{sigmund2010social}
Karl Sigmund, Hannelore De~Silva, Arne Traulsen, and Christoph Hauert.
\newblock Social learning promotes institutions for governing the commons.
\newblock {\em Nature}, 466(7308):861--863, 2010.

\bibitem{traulsen2006stochastic}
Arne Traulsen, Martin~A Nowak, and Jorge~M Pacheco.
\newblock Stochastic dynamics of invasion and fixation.
\newblock {\em Physical Review E—Statistical, Nonlinear, and Soft Matter Physics}, 74(1):011909, 2006.

\bibitem{nowak2004emergence}
Martin~A Nowak, Akira Sasaki, Christine Taylor, and Drew Fudenberg.
\newblock Emergence of cooperation and evolutionary stability in finite populations.
\newblock {\em Nature}, 428(6983):646--650, 2004.

\bibitem{imhof2005evolutionary}
Lorens~A Imhof, Drew Fudenberg, and Martin~A Nowak.
\newblock Evolutionary cycles of cooperation and defection.
\newblock {\em Proceedings of the National Academy of Sciences}, 102(31):10797--10800, 2005.

\bibitem{shen2025mutation}
Chen Shen, Zhixue He, Lei Shi, and Jun Tanimoto.
\newblock Mutation mitigates finite-size effects in spatial evolutionary games.
\newblock {\em Communications Physics}, 8(1):1--8, 2025.

\bibitem{nowak1992evolutionary}
Martin~A Nowak and Robert~M May.
\newblock Evolutionary games and spatial chaos.
\newblock {\em nature}, 359(6398):826--829, 1992.

\bibitem{dafoe2020open}
Allan Dafoe, Edward Hughes, Yoram Bachrach, Tantum Collins, Kevin~R McKee, Joel~Z Leibo, Kate Larson, and Thore Graepel.
\newblock Open problems in cooperative ai.
\newblock {\em arXiv preprint arXiv:2012.08630}, 2020.

\bibitem{nowak2006five}
Martin~A Nowak.
\newblock Five rules for the evolution of cooperation.
\newblock {\em science}, 314(5805):1560--1563, 2006.

\bibitem{han2022institutional}
The~Anh Han.
\newblock Institutional incentives for the evolution of committed cooperation: ensuring participation is as important as enhancing compliance.
\newblock {\em Journal of The Royal Society Interface}, 19(188):20220036, 2022.

\end{thebibliography}

\end{document}